# Millimeter-Wave Spectroscopy of the Organic spin-Peierls System β′-(ET)$_2$SF$_5$CF$_2$SO$_3$

Brian H. Ward, Isaac B. Rutel, James S. Brooks[*]
*NHMFL-Florida State University*
*Tallahassee, Florida 32310*

John A. Schlueter
*Chemistry and Materials Science Divisions, Argonne National Laboratory*
*Argonne, Illinois 60439-4831*

R. W. Winter and Gary L. Gard
*Department of Chemistry, Portland State University*
*Portland, Oregon 97207-0751*

[*]To whom correspondence should be addressed.

**Abstract**

The first *purely* organic BEDT-TTF spin-Peierls system, $\beta'$-(ET)$_2$SF$_5$CF$_2$SO$_3$, has been confirmed using a high-frequency electron spin resonance (EPR) cavity perturbation technique. The material exhibits the characteristics of a quasi-one-dimensional (1D) Heisenberg antiferromagnetic spin system above 30 K, but undergoes a second-order transition, at $T_{SP}$ = 33 K, to a singlet ground state, due to a progressive spin-lattice dimerization. The spin-Peierls state is evidenced by a sharp drop in the spin susceptibility below 24 K for the magnetic field parallel to each of the three principle axes (i.e. $H$ $a$, $H$ $b$, and $H$ $c$). The 1D chain axis has been identified as the crystallographic *b* axis from the *g* value analysis. The singlet-triplet gap, $\Delta_\sigma(0)$ = 114 ( 21) K, was determined using a modified BCS theory. Also, we describe in some detail the millimeter-wave vector network analyzer (MVNA) for researchers who have interest in precision EPR measurements at higher magnetic fields and frequencies.

## 1. Introduction

There has been much interest in the *bis*(ethylenedithio)tetrathiafulvalene (BEDT-TTF or ET) radical cation salts since the discovery of the $(ET)_2ReO_4$ superconductor in 1983.[1] This led to the synthesis of compounds, which adhere to a $(ET)_2X$ stoichiometry, allowing for the formation of several compounds with varying physical, electronic, and magnetic structures.[2]  Among these salts, $\kappa\text{-}(ET)_2CuN(CN)_2Br$ has the highest superconducting transition temperature at ambient pressure, $T_c = 11.6$ K.[3]

Further advancement was achieved when Geiser et al.[4] reported on $\beta''\text{-}(ET)_2SF_5CH_2CF_2SO_3$, which goes through a superconducting transition at 5.2 K, in 1996. This report was interesting due to the use of the completely organic anion $SF_5CH_2CF_2SO_3^-$. Further systematic investigation led to the synthesis of several other organic anions, $SF_5CH_2SO_3^-$, $SF_5CHFSO_3^-$, $SF_5CF_2SO_3^-$; and subsequently their corresponding $(ET)_2X$ salts.[5]

Small differences in the anion, have led to large differences in the physical properties among the three new organic salts. For instance $\beta'' \text{-} (ET)_2SF_5CH_2SO_3$ is semiconducting over a range of temperature from 4.2 K to 300 K, while $\beta'' \text{-} (ET)_2SF_5CHFSO_3$ is semiconducting from 100 – 300 K, metallic from 6 – 100 K, and undergoes a metal-to-semiconductor transition below 10 K.[5]  Finally, $\beta'\text{-} (ET)_2SF_5CF_2SO_3$ shows semiconducting behavior above the antiferromagnetic ordering transition temperature of around 20 – 24 K.[5]

β′-(ET)$_2$SF$_5$CF$_2$SO$_3$, like most other 2:1 salts of ET, has layers of donor molecules separated by layers of the anions. Salts of the organic donor molecule ET have a variety of crystal packing morphologies, which dictate the physical properties of the salts. In β′-(ET)$_2$SF$_5$CF$_2$SO$_3$ the packing morphology is designated β′ because the cation-radical salt forms strongly dimerized, face-to-face ET molecules with the ET dimers along the stacking axis displaced from one another. The β′-phase materials are poorly conducting insulators, at all temperatures, with highly one-dimensional (1D) energy band structures.[6] The ground state of these 1D systems is a half-filled band and they are insulating due to large Coulomb repulsion. There are several other examples of β′-(ET)$_2$X salts, namely β′-(ET)$_2$ICl$_2$, β′-(ET)$_2$BrICl, and β′-(ET)$_2$AuCl$_2$.[7,8] All of the β′-phase materials are low-dimensional, localized spin systems, as exhibited by their Bonner-Fisher (BF)[9] or quadratic layer Heisenberg antiferromagnet (QLAF)[10] type magnetic susceptibility. Typical temperature-dependent susceptibility data contains a broad maximum at high temperatures followed by a transition to an antiferromagnetic ground state at low temperatures (T < 30 K).[7,8]

In 1997 Yoneyama et al.[11] reported the single crystal magnetic susceptibility and electron paramagnetic resonance (EPR) data for the β′-(ET)$_2$ICl$_2$, and β′-(ET)$_2$AuCl$_2$ salts. Their data show that these two molecular conductors are two-dimensional (2D) layered Heisenberg antiferromagnets which undergo three-dimensional (3D) antiferromagnetic ordering at $T_N$ = 22 K and 28 K for the ICl$_2^-$ salt and the AuCl$_2^-$ salt, respectively.

Other examples of organic antiferromagnetic systems, namely spin-Peierls systems, are limited to (TTF)$MX_4C_4(CF_3)_4$ ($M$ = Cu, Au; $X$ = S, Se), MEM(TCNQ), (BCPTTF)$_2$X ($X$ = PF$_6$, AsF$_6$), and (TMTTF)$_2$PF$_6$.[12-15] The spin-Peierls transition temperatures for the listed salts range from 2 to 36 K. Also in 1989, Obertelli et al. reported evidence for the first ET-based spin-Peierls system, α'-(ET)$_2$Ag(CN)$_2$.[16] The transition temperature in the Ag(CN)$_2$ salt was ~7 K.

Herein we present an orientation dependent high-frequency (V-band) EPR study which unambiguously demonstrates that β'-(ET)$_2$SF$_5$CF$_2$SO$_3$ is the first *purely* organic ET-based spin-Peierls system at $T_{SP}$ = 33( 7) K. We demonstrate, through the lack of thermal hysteresis, the second-order nature of the transition. Also, using a BCS-like theory, we calculate a value for the spin-gap, $\Delta_\sigma(0)$, of 114( 21) K. The singlet-ground state forms via dimerization of the spin chains induced by spin-phonon coupling.

## 2. Experimental Section

*Sample Preparation.* Crystals of the β'-(ET)$_2$SF$_5$CF$_2$SO$_3$ salt were grown by electrocrystallization methods previously described.[5] The crystal utilized in this experiment was a shiny, black rod with the dimensions 2.0 x 0.4 x 0.4 mm$^3$.

*Instrumentation and Design.* The EPR data were collected using a millimeter-wave vector network analyzer (MVNA manufactured by AB*mm*, 52 Rue Lhomond, 75005 Paris, France).[17] The sample was placed into a cylindrical, copper cavity with silicone grease, halfway between the cavity's axis and its wall. The cavity has an inner radius of 9.55 cm and a length of 9.31 cm where the predominate modes excited are the TE01$p$ ($p$ = 1, 2,…) modes. A heating coil attaches to the bottom of the cavity using one of the mounting screws. The cavity couples to the wave guides through two isolated holes 1.65 mm in diameter. The holes are positioned on the coupling plate by centering them in the cross sectional area of the wave guides while also maintaining equidistance from the axis of the cavity. The coupling plate then attaches to the two stainless steel wave guides, which are rectangular and lead to the vacuum tight windows. These windows allow the electromagnetic radiation through, but maintain a vacuum seal inside the jacket. The windows connect to the diode mounting plates by short lengths of more wave guide material.

The Schottky diodes are placed on the mounting plates and connected to the MVNA through high-frequency coaxial cable. The V diodes used in our experiment allowed for frequencies between 48 – 70 GHz, and the current investigation focused primarily on the 67.5 GHz resonance. The frequency was held steady by an EIP Model 578B Source Locking Microwave Counter, and can be controlled by the MVNA program run on a Hewlett Packard Vectra VL Series 3 computer. The coax from the diodes is connected the vector analyzer which is then connected to a HAMEG 30MHz Oscilloscope. A Stanford Research Systems (SRS) SR830 DSP Lock-In Amplifier monitored the output from the oscilloscope at a frequency of 488.2 Hz.

A 9 Tesla superconducting magnet powered by a CRYOMAGNETICS IPS-100 Magnet Power Supply, and a CRS-100 Current Reversing Switch provided the magnetic field. To monitor the magnetic field, the voltage across the magnet was measured with a Keithley digital multimeter, while the temperature of the probe was controlled using a Conductus LTC-20 temperature controller.

The lock-in amplifier, digital multimeter, and temperature controller levels were recorded using LabView 5.0 on a Power Macintosh 7100/66. The final analysis was performed using Igor Pro 3.12 from WaveMetrics Inc. on a Dell Optiplex GX1p running Windows NT.

**3. Results**
*Orientation Dependent EPR Spectra.* Performing orientation dependent EPR experiments can identify whether a material is a 3D antiferrromagnetically ordered system or a non-magnetic spin-Peierls system. The susceptibility for a 3D antiferromagnetically ordered system vanishes for only one crystal orientation, while in the spin-Peierls system the susceptibility disappears for all three principle crystal orientations. In three separate experiments a rod-like crystal (the same sample for all three runs) of β′-(ET)$_2$SF$_5$CF$_2$SO$_3$ was mounted with each of the three principle axes parallel to the magnetic field, (i.e. $H \parallel a$, $H \parallel b$, and $H \parallel c$). Figure 1 shows the packing pattern of the ET donor molecules in β′-(ET)$_2$SF$_5$CF$_2$SO$_3$.[5] The dotted lines between molecules indicate intermolecular contacts less than 3.6 Å (i.e., the sum of the sulfur van der Waals radii). Figure 2 plots the evolution of the EPR line for β′-(ET)$_2$SF$_5$CF$_2$SO$_3$ along each of the three principle crystallographic axes. The linewidth, *ΔH*, of the EPR resonance, defined as the full width at half maximum (FWHM), ranges from 0.2, 0.6, and 2.0 mT for $H \parallel c$, $H \parallel a$, and $H \parallel b$, respectively.

*Spin Susceptibility.* Figure 3 shows the temperature dependence of the EPR spin intensity of β′-(ET)$_2$SF$_5$CF$_2$SO$_3$ for all three orientations. Upon lowering the temperature, the relative spin susceptibility in all three orientations increases slightly from 50 K to a maximum at 25 K. Below 25 K the spin susceptibility drops precipitously to zero. The spin intensity for $H \parallel c$ was fit to an absorption probability function for the singlet ground state as well as a modified BCS spin-gap function.[18] From the absorption probability fit, shown in Figure 3, the values for $\Delta_\sigma$ and $T_{SP}$ are 125.85 K and 35.5 K, respectively. From an Arrhenius-type plot the values for $\Delta_\sigma$ and $T_{SP}$ from BCS theory are 114 ($\pm$ 21) K and 33 ($\pm$ 7) K, respectively.

*g* **Value Anisotropy Analysis.** The temperature dependence of the *g* value is plotted in Figure 4 from 50 to 18 K for the $H \parallel c$ and $H \parallel b$ orientations and from 50 to 26 K for the $H \parallel a$ orientation. The *g* values for $H \parallel a$ could only be calculated down to 26 K, at which point the signal could not be resolved due to the overlap with the DPPH signal. Below 18 K for $H \parallel c$ and $H \parallel b$ the EPR signal is non-existent. The *g* value is nearly temperature independent for the $H \parallel a$ and $H \parallel c$ orientations. However, the *g* value increases slightly, for the $H \parallel b$ orientation, from approximately 2.003 to 2.006 through the transition.

**4. Discussion**

Orientation dependent EPR measurements of organic conductors can provide information on spin susceptibility, $\chi_s$, and *g* value anisotropy that is related to the crystallographic and magnetic structure.[19] As previously shown,[5] $\beta'$-(ET)$_2$SF$_5$CF$_2$SO$_3$ has layers of donor molecules separated by layers of charge compensating anions. The unique feature of the $\beta'$-(ET)$_2$SF$_5$CF$_2$SO$_3$ salt is the strongly dimerized ET molecules within a stack. Typically, the $\beta'$-phase ET salts (i.e. $\beta'$-(ET)$_2$ICl$_2$, and $\beta'$-(ET)$_2$AuCl$_2$)[7,8] are layered Heisenberg antiferromagnets that experience 3D antiferromagnetic ordering at low temperatures. Contrary to $\beta'$-(ET)$_2$ICl$_2$, and $\beta'$-(ET)$_2$AuCl$_2$, $\beta'$-(ET)$_2$SF$_5$CF$_2$SO$_3$, has a transition to a spin-Peierls non-magnetic ground state which is seen clearly in the orientation dependent EPR data.

*EPR Spectra Anaylsis.* The EPR spectra displayed in Figure 2 were collected between 10 and 50 K. Resonance anisotropy is typical in low-dimensional molecular solids. The movement of the ET resonance in β′-(ET)$_2$SF$_5$CF$_2$SO$_3$ can be seen in Figure 2. Because of slight misalignments of our probe from the center of field (COF), a DPPH marker was used as a calibration and appears as a single absorption up field from the ET signal. The intensity of the EPR line for $H \parallel a$, and $H \parallel c$ is noticeably greater than for $H \parallel b$. The differences can be explained by examining the crystal structure. Figure 1 shows that the ET molecules form dimerized stacks approximately along the *a* axis, while sheets are formed in the *bc* plane. Based on close S⋯S contacts, there are few *intrastack* interactions between dimers and numerous *interstack* interactions. When the crystal is oriented with $H \parallel b$, the excited AC magnetic field currents actually run parallel to the ET stacks (i.e. the *a* axis). Magnetic exchange is minimal along the stacks producing a weak EPR signal. In both the $H \parallel a$, and $H \parallel c$ orientations the *bc* plane is parallel to the AC magnetic field currents. Magnetic exchange is stronger in the *bc* plane resulting in a more intense EPR signal.

Each EPR spectrum was fit to a Lorentzian line shape. The ET signals for the $H \parallel b$ and $H \parallel c$ orientations are well separated from the DPPH line, but for the $H \parallel a$ orientation the ET and DPPH signals overlap. In the latter case the data were fit using a double Lorentzian line shape function. From the integrated signals of Figure 2 for each orientation we obtain the EPR intensity as a function of temperature (Figure 3). In EPR experiments the signal intensity is a direct measure of the spin susceptibility. The abrupt drop in the spin intensity in all three orientations reflects the transition to a non-magnetic singlet ground state, and the opening of a finite energy gap in the excitation spectrum.[13] The opening of a spin gap is indicative of a spin-Peierls transition. The spin-Peierls transition is driven by 1D antiferromagnetic fluctuations that couple to the lattice through spin-phonon interactions.[20]

The narrow EPR lines observed in organic conductors are attributed to the low-dimensionality of the materials, which reduces the spin-orbit coupling. The observed $\Delta H$, values agree with those calculated from the X-band EPR data.

Previously reported X-band EPR data for $\beta'$-(ET)$_2$SF$_5$CF$_2$SO$_3$ can be described using the well-known BF model for a 1D antiferromagnetic Heisenberg chain system. The data show the characteristic broad maximum above the transition temperature indicative of short-range antiferromagnetic coupling in low-dimensional systems, while the fit yields a magnetic exchange constant, $J_{BF}$, of 257 K. The X-band EPR data also show a steep drop in intensity for all three orientations below the spin-Peierls transition temperature.

Our spin-Peierls transition temperature, $T_{SP}$, using an operating frequency of 62.58 GHz (V-band) is lower than the value determined by X-band EPR. The source of this discrepancy is the differing operating external magnetic fields of the two experiments. In the V-band EPR experiment the magnetic field is swept between 2.45 and 2.48 T, while in the X-band experiment the field is swept between 0.3 and 0.36 T. The response of a spin-Peierls system to an external magnetic field, $H$, is a monotonically decreasing $T_{SP}$ as $H$ is increased. This behavior can be understood in terms of quantum fluctuations, which provide the driving force for the spin-Peierls transition.[13] Because $H$ reduces the quantum fluctuations, it reduces the energy available to form the spin-Peierls phase resulting in a lower value of $T_{SP}$. Using the value of $T_{SP}$ from our fit as the zero-field point, as well as $T_{SP}$ determined from data collected at 9.25, 67.5, and 82.3 GHz, we have constructed a preliminary phase diagram for $\beta'$-(ET)$_2$SF$_5$CF$_2$SO$_3$ (inset of Figure 3). Our data are consistent with a monotonically decreasing spin-Peierls transition temperature as the magnetic field is increased.

We also performed a thermal hysteresis experiment through the spin-Peierls transition for the $H$ $c$ orientation. EPR data were collected from 12 – 20 K in 1 K increments for both cooling and warming runs. The lack of thermal hysteresis through the transition signifies a second-order phase transition, which agrees with spin-Peierls theory.[13]

From previous $g$ value analysis on the spin-Peierls system (TMTTF)$_2$PF$_6$,[15,21] we believe that the increase in $g$ value below 25 K identifies the $b$ axis as the spin-Peierls quasi-1D chain axis. In other words, the quasi-1D chains are formed through *inter*stack (i.e. the $b$ axis) ET dimer interactions rather than *intra*stack interactions (i.e. the $a$ axis). Yoneyama et al.[11] reported the orientation-dependent EPR and magnetic susceptibility data for the $\beta'$-(ET)$_2$ICl$_2$, and $\beta'$-(ET)$_2$AuCl$_2$ salts. Their data suggests that the ICl$_2$ and the AuCl$_2$ salts undergo 3D antiferromagnetic ordering below $T_N$ = 22 K and 28 K, respectively. They also point out that the magnetic easy axis in the antiferromagnetically ordered ICl$_2$ and AuCl$_2$ salts is the crystallographic $c$ axis. Due to differing crystal coordinate systems, the $c$ axis in the $\beta'$-(ET)$_2$ICl$_2$, and $\beta'$-(ET)$_2$AuCl$_2$ salts corresponds to the $b$ axis in $\beta'$-(ET)$_2$SF$_5$CF$_2$SO$_3$. Therefore, the quasi-1D ET chain axis is the same for both the 3D antiferromagnetically ordered systems as well as our spin-Peierls system.

The 1D structural fluctuations in $\beta'$-(ET)$_2$SF$_5$CF$_2$SO$_3$ along the 1D ET chains (i.e. $b$ axis ) couple to the lattice via spin-phonon coupling. Below $T_{SP}$ the structure of the underlying magnetic lattice changes producing dimerized chains. The 1D fluctuations and the chain dimerization associated with the spin-Peierls state effectively reduce spin-orbit coupling along the *b* axis direction resulting in a *g* value shift as a function of temperature.[13]

Finally, the value for the spin-Peierls gap in β′-(ET)$_2$SF$_5$CF$_2$SO$_3$ was analyzed within a BCS-like mean-field model,[18] as well as a temperature-dependent (i.e. below the gap) absorption probability function for the singlet ground state.[22] The BCS-like mean-field results were a transition temperature of 33 ( 7) K and a gap value of 114 ( 21) K based on several fits. The calculated values for the transition temperature and gap energy were 35.5 K and 125.85 K, respectively, using the absorption probability function for the singlet ground state. The temperature independent BCS-like function and the temperature-dependent absorption probability function give similar values for $T_{SP}$ and $\Delta_\sigma$ within acceptable error limits.

*The MVNA Instrument and Resonant-Cavity-Based Measurements.* To investigate the magnetic properties described above we employed a relatively new technique of non-modulated high-frequency resonant perturbation detection. To monitor the signal in a resonant cavity, we use a millimeter-wave vector network analyzer (MVNA). The signal contains both the phase and amplitude of the millimeter-wave radiation transmitted through the resonant cavity. Both the MVNA instrument and the resonant cavity design will be briefly discussed below.

The MVNA operates at high frequencies by using a non-linear device, a Schottky diode, to multiply, and then modulate the frequency by integer multiples. The frequency multiplier is also referred to as the Harmonic Generator (HG) and produces frequencies in the millimeter range. The chosen harmonic of the HG is then paired to the appropriate Harmonic Modulator (HM) to allow for detection of the signal. The system is based on two Yttrium-Iron-Garnet (YIG) sources, which are continually tunable from 8 – 18 GHz. Using pairs of Schottky diodes it is possible to generate frequencies ranging from $N*(8 – 18$ GHz), where $N$ is an integer from 3 to 20, or 8 to ~200 GHz.[23]

The use of a resonant cavity offers many advantages in the millimeter frequency range. In the case of particularly small crystals, the radiation wavelength may be large compared to the sample dimensions. Due to the high $Q$-factor of the resonance, small changes in the sample lead to large changes in the electromagnetic field response, also known as the "cavity perturbation" technique.[17,23]

Using the strongly enhanced sensitivity of this method, $Q$-factors ranging from 5000 to 20000 can be achieved. The sensitivity is enhanced over a single-pass measurement for samples of comparable size by the $Q$-factor, as a multiplier (approximately).[17] This is critical for accurate measurements when dealing with small single crystal samples, and equally small signals. The resonant cavity utilized in our experiment, shown in Figure 5, is cylindrical and made of copper. It allows for the excitation of many modes, however, the TE01$p$ ($p = 1,2,$ ) modes (Figure 5) are excited with the highest $Q$-factors. Using this setup, the DC magnetic field is applied parallel to the end plate of the cavity allowing a large range of frequencies to be employed, without the need of a sample holder. The sample is placed halfway between the center of the cavity and the cavity wall to ensure that it is sitting within the radial AC magnetic fields associated with the TE01$p$ modes of the cavity.[17] Deviation from the midpoint greatly reduces the sensitivity of the measurement.

The MVNA system is a versatile set-up that allows for investigation of several high-frequency measurements with little modification. In our experiments, we utilize the reflective characteristics of the resonant system, but the MVNA can be quickly modified for transmission type experiments, and for investigations at other limited frequencies above 200 GHz (up to ~700 GHz by association with Gunn oscillators).[17] The MVNA provides greater sensitivity and better resolution of magnetic transitions than traditional X-band EPR, since higher frequencies excite transitions at higher (more separated) fields. Evidence for the high sensitivity and resolution of the MVNA instrument is seen in the clarity of the β′-(ET)$_2$SF$_5$CF$_2$SO$_3$ and the DPPH resonances, which range from 0.01 – 0.1 mT. Also, the style of the probe permits geometric versatility in cavity design, allowing more freedom to access the type of modes a researcher wishes to investigate, with minimal modification. The MVNA system is adaptable to many types of experiments, and prompts further investigation as a research aid.

## 5. Conclusions

The first *purely* organic BEDT-TTF spin-Peierls system, β′-(ET)$_2$SF$_5$CF$_2$SO$_3$, has been confirmed using a high-frequency electron spin resonance (EPR) cavity perturbation technique. The material exhibits the characteristics of a quasi-one-dimensional (1D) Heisenberg antiferromagnetic spin system above 30 K, but undergoes a second-order transition, at $T_{SP} = 33$ K, to a singlet ground state, due to a progressive spin-lattice dimerization. The lack of thermal hysteresis through the transition signifies a second-order phase transition, which agrees with spin-Peierls theory. The spin-Peierls state is evidenced by a sharp drop in the spin susceptibility below 24 K for the magnetic field parallel to each of the three principle axes (i.e. $H \parallel a$, $H \parallel b$, and $H \parallel c$). The singlet-triplet gap, $\Delta_\sigma(0) = 114 \ (\pm 21)$ K, was determined using a modified BCS theory. The spin-Peierls transition is driven by 1D antiferromagnetic fluctuations that couple to the lattice through spin-phonon interactions. We describe in some detail a relatively new technique of non-modulated high-frequency resonant perturbation detection using a millimeter-wave vector network analyzer (MVNA). The MVNA system is a versatile set-up that allows for investigation of several high-frequency measurements (40 – 200 GHz) with little modification. The MVNA provides greater sensitivity and better resolution of magnetic transitions than traditional X-band EPR, since higher frequencies excite transitions at higher (more separated) fields. The extreme sensitivity of the MVNA technique makes it an excellent system for performing EPR investigations.

**Acknowledgment.** Work at NHMFL-FSU was supported by NSF Grant No.DMR-99-71474 and NHMFL/IHRP 500/5031. The NHMFL is supported through a contractual agreement between the NSF through Grant No. NSF-DMR-95-27035 and the State of Florida. Work at Argonne National Laboratory was supported by the US Department of Energy, Office of Basic Energy Sciences, Division of Materials Sciences, under contract No. W-31-109-ENG-38. Work at Portland State University was supported by NSF Grant No. CHE-9904316 and the Petroleum Research Fund ACS-PRF 34624-AC7.

**References.**

**Figure Captions.**

**Figure 1.** View of the ET cation layer in β′-(ET)$_2$SF$_5$CF$_2$SO$_3$ looking down the long axis of the molecule. The dashed lines indicate S⋯S van der Waals contacts less than 3.6 . The ET molecules are dimerized and form 1D chains along the *b* axis.

**Figure 2.** Evolution of the EPR signal as the temperature is raised from 10 to 50 K. (Spectra are offset for clarity.) DPPH, included for field calibration, appears as a single resonance up field from the ET resonance. Each spectrum was fit to a Lorentzian line shape.

**Figure 3.** The temperature dependence of the EPR integrated absorption for the resonance line at 67.5 GHz for all three orientations ($H$  $a$, $H$  $b$, and $H$  $c$). The inset shows the fit (solid line) below $T_{SP}$ using equation 1.

**Figure 4.** The temperature dependence of the *g* value for all three orientations. For the *a* and *c* axis parallel to the external magnetic field the *g* value is relatively constant, while for the *b* axis parallel to the external magnetic field the *g* value increases below $T_{SP}$.

**Figure 5.** A schematic of the experimental setup, including the Schottky diodes, wave guides, coupling plate, cavity, vacuum jacket, and the magnet. The dashed line indicates the center of the magnet. A 3D view of the cavity showing the position of the sample at the bottom of the cavity (upper part). The AC magnetic field ($H_{AC}$) distribution within the cavity for the TE011 mode. The external DC magnetic field ($H_{DC}$) is applied parallel to the cavity axis so that $H_{AC}$ is perpendicular to $H_{DC}$ (lower part).

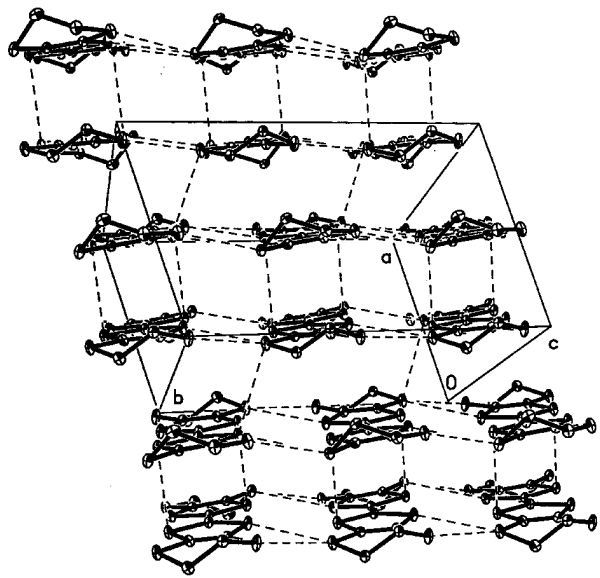

**Figure 1.**

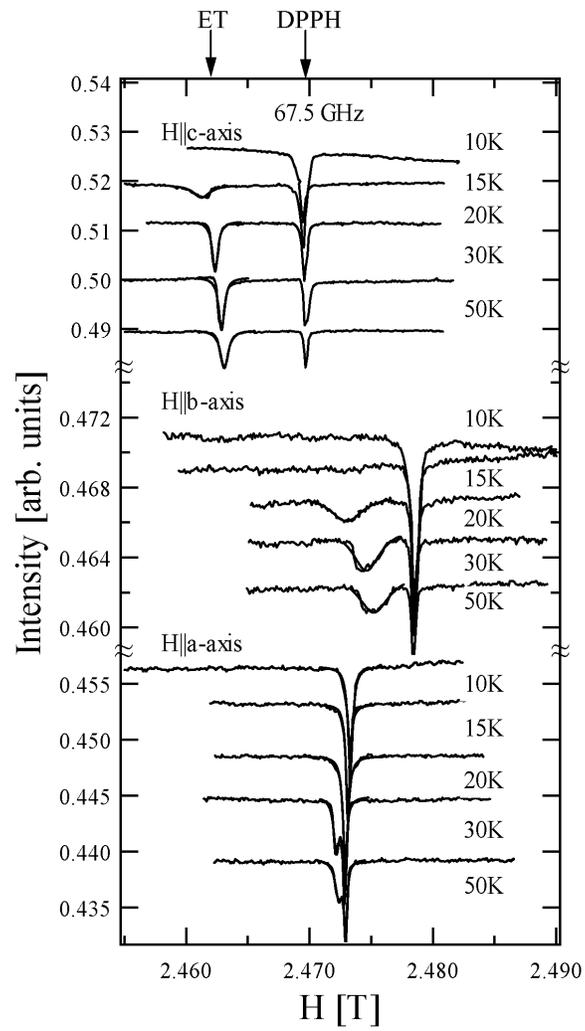

**Figure 2.**

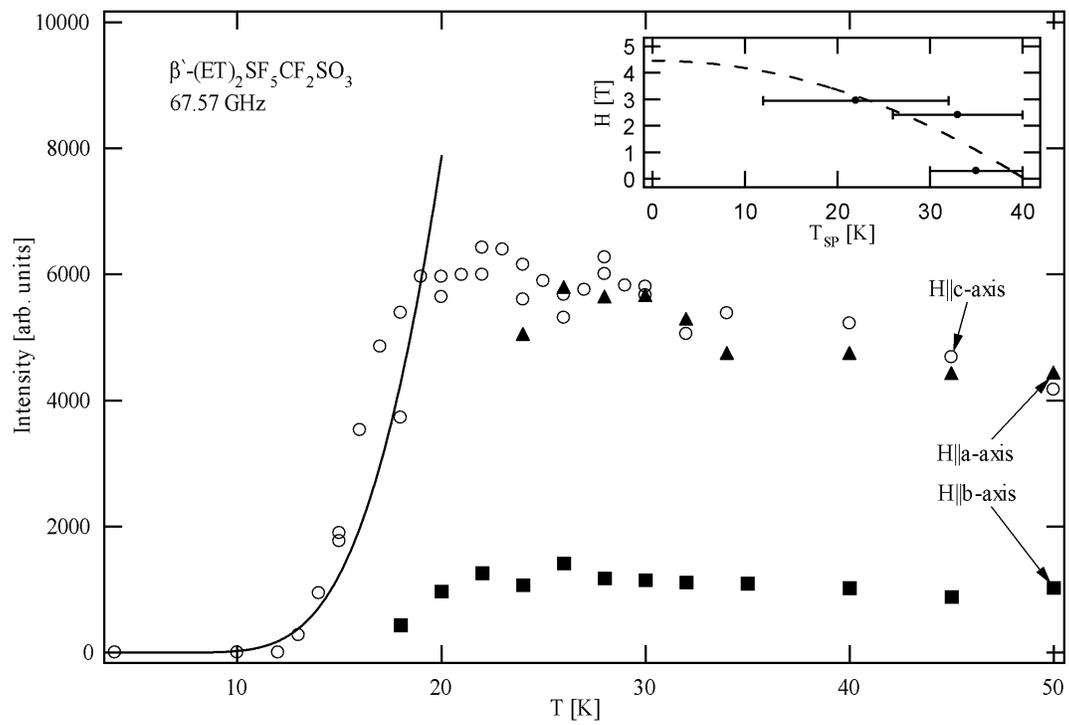

**Figure 3.**

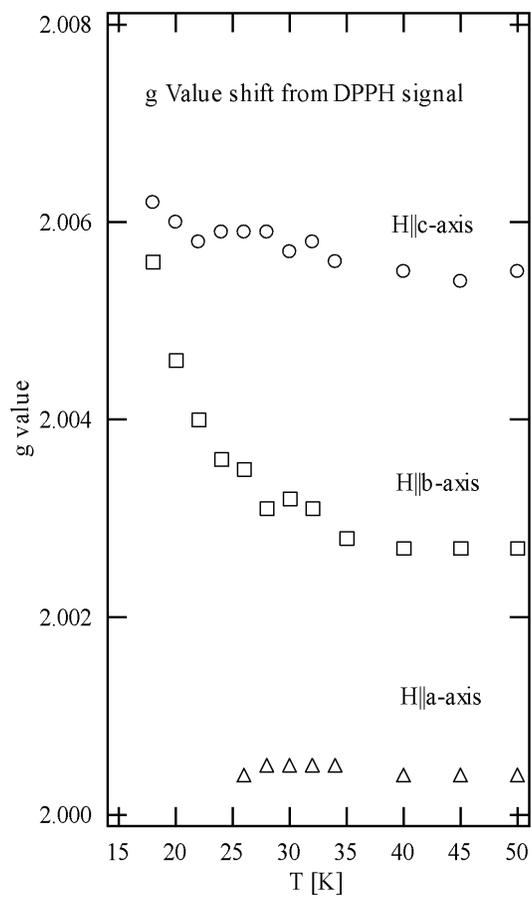

**Figure 4.**

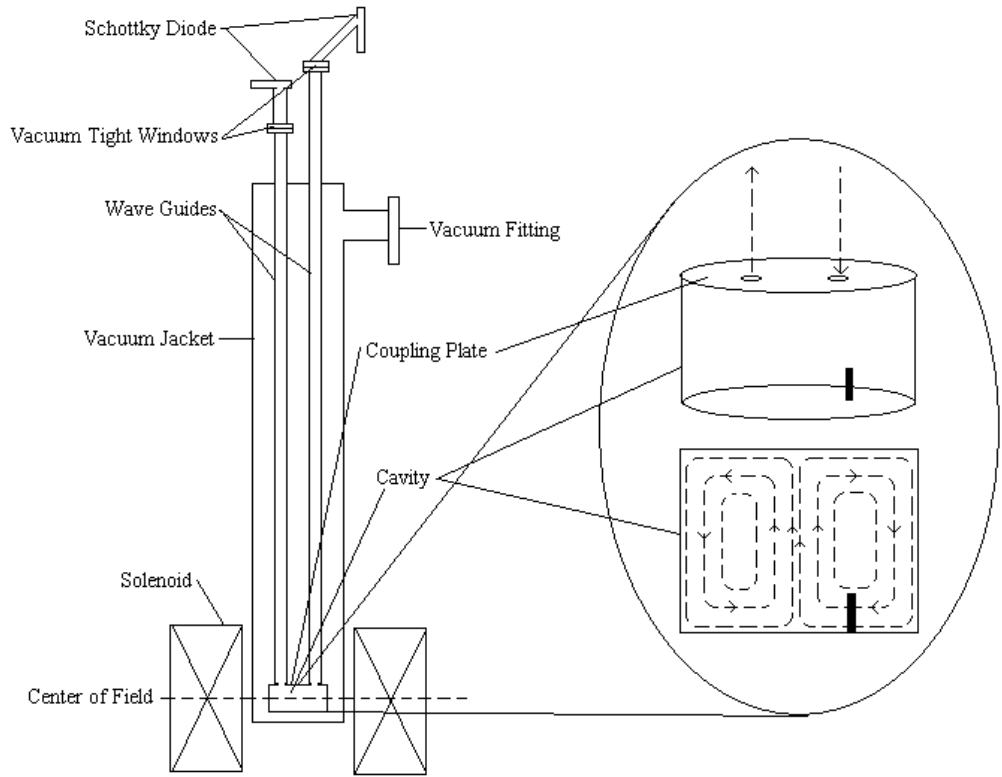

**Figure 5.**